\documentclass[%
 reprint,
 amsmath,amssymb,
 aps,pre,
floatfix
]{revtex4-1}

\usepackage[dvipdfmx]{graphicx}
\usepackage{dcolumn}
\usepackage{bm}

\begin{document}


\title{Exact expression for average kinetic energy  \\ of 2-dimensional freely jointed chain and a related model}

\author{Tetsuro Konishi}
 \affiliation{College of Engineering, Chubu University, Kasugai, 487-8501,Japan.}
\author{Tatsuo Yanagita}%
\affiliation{%
Department of Engineering Science,  Osaka  Electro-Communication University, Neyagawa, 572-8530, Japan.
}%




\date{\today}

\begin{abstract}
For a 2-dimensional freely jointed chain with 3 particles and a related model,
the average and variance of the kinetic energies of each particle in thermal equilibrium  are exactly obtained. The same is done for a related model.
The excess of average kinetic energies near the chain ends, previously observed by numerical simulation, is analytically confirmed.
The non-uniformity of the average kinetic energy results from the generalized principle of equipartition of energy.
%
The non-uniformity also depends  on temperature for a model that has  intra-chain potential, and we can control it from outer-energetic to inner-energetic by decreasing the temperature.

\begin{description}
\item[PACS numbers]
\end{description}
\end{abstract}

\pacs{Valid PACS appear here}
\maketitle


\section{Introduction}

When a material is at some finite temperature, its components, for instance,  molecules and atoms, show random motion, and the magnitude of this random motion is measured by the kinetic energy of each components.  The kinetic energy  of each component is often proportional to the temperature, which is guaranteed by the principle of equipartition of energy.

The principle of  equipartition of energy is a fundamental property in classical statistical mechanics, and is  satisfied by  systems in thermal equilibrium~\cite{doi:10.1098/rspl.1843.0077,doi:10.1098/rstl.1867.0004,Tolman-PR-1918,Tolman-book,kubo-book}.  
The principle states that if a system  is in thermal equilibrium, 
the  average   kinetic energy is the same for  every  degree of freedom, and is proportional to the temperature.
That is,   the principle guarantees that the average kinetic energy is {\sl uniform} over the entire system.
This principle  can be derived  when the  kinetic energy of the system is expressed by
sum of squared momentuma with coefficients.

This principle is utilized in various ways.  For example,  one can determine the temperature of the system by measuring the kinetic energy.
It  can also be  used  as  a measure of relaxation to thermal equilibrium~\cite{ergodic-measure-kirkpatric,ergodic-measure-straub,sheet-tkg-1,sheet-tgk-2,sheet-tgk-3}.
The principle states that if the system is in thermal equilibrium, the average kinetic energy is uniform. Therefore,  if the average kinetic energy is not uniform, the system is not in thermal equilibrium.  
For example, in the well-known Fermi-Pasta-Ulam problem~\cite{FPU,Dauxois-2005,CHAOS-AIP-FPU50},  they numerically integrated the equation of motion for non-linear lattice model and measured   the energy of modes.
They found that the modes do not have the same energy, and thus, even though the model has anharmonic interaction between degrees of freedom, and energy can flow from one mode to another, the system does not show ergodic behavior for finite time.


Although the principle of equipartition of energy is a common feature for systems in thermal equilibrium,  some systems  do not obey this principle.
In 1998 one of the authors numerically found that the average kinetic energies of the masses in a multiple pendulum were not equal: the masses  near the root had small values and the ones near the end had large values~\cite{yanagita-gakkai-1,yanagita-gakkai-2,yanagita-gakkai-3}. 

Non-uniformity of average kinetic energy is also found by the authors  for freely jointed chain, which is a system composed of masses connected by rigid links~\cite{chain-letter-JSTAT-2009}. In this model,  the  average kinetic energy is larger near the chain ends.
Both of these systems, i.e., the multiple pendulum and the freely jointed chain,  have a property in common 
that the kinetic energies of the systems have off-diagonal elements which depend on 
coordinates.


Since the principle of equipartition of energy assumes that the kinetic energy is expressed as
the sum of squared momenta,  the principle does not hold for  systems
whose kinetic energy is expressed as a quadratic form of momenta with off-diagonal elements.
For such systems we have the ``generalized principle of equipartition of energy''~\cite{Tolman-PR-1918,Tolman-book,kubo-book} , where averages of other quantities, different from kinetic energy of each degree of freedom, takes the same value proportional to temperature.
The multiple pendulum (including the double pendulum) and the freely jointed chain do not satisfy the assumption of the principle of equipartition but satisfy that of  the generalized principle of equipartition. This is because these systems have off-diagonal elements in the kinetic energy, that are induced by holonomic constraings. In fact, for the above mentioned cases the constraints are that the particles are connected by rigid links, and the distances between adjacent particles are kept fixed.


Note that, for systems for which  the generalized principle of equipartition holds but the conventional principle does not, the average kinetic energy of  each particle is not uniform.
Now we arrive at an important property: {\sl for systems with holonomic constraints, kinetic energy of each particle can be non-uniform in thermal equilibrium}.
If we regard the average kinetic energy as proportional to the temperature,  the non-uniformity of average kinetic energy implies the non-uniformity of ``local temperature'' at thermal equilibrium.

Although many years have passed since the generalized principle of equipartition was established,
its implication has'nt been investigated throughly.
It will be interesting to find some systematic non-uniformity in the energy distribution at thermal equilibrium, particularly for functional  molecules such as proteins.
In the paper~\cite{chain-letter-JSTAT-2009} we studied  the freely jointed chain~\cite{fjc-Kuhn-1934,KUHN194811,Kramers-chain,Fixman3050,doi:10.1063/1.1681834,Mazers-chain-pre-1996,doi-edwards,doi-intro-polymer,strobl-polymer,SiegertWinklerReineker+1993+584+594}. In this model, adjacent masses are serially connected by rigid links. Thus, it  has holonomic constraints and is subject to the generalized principle of equipartition of energy. 
Using numerical integration of the equation of motion and approximate analytic calculation we found that, if the mass of each particle is equal,  the average kinetic energy has larger  values near the chain ends  than  in the middle of the chain. 
From the generalized principle of equipartition of energy it is plausible that the value of the average kinetic energy is not uniform but  depends on the position. The results show that how the average kinetic energy systematically deviates from uniformity.

As mentioned above, systems with holonomic constraints  usually do not obey the conventional principle of equipartition and show spontaneous non-uniformity for the average kinetic energies of each degree of freedom.
For such systems, it will be valuable to obtain an analytic expression of the average kinetic energy.

In this paper we present  two models in which we can obtain exact analytic expressions for the kinetic energy of each particle in thermal equilibrium.
%
The  models discussed include  a 2-dimensional freely jointed chain with three particles and another  modified model.
The results  described in this paper reveal the spontaneous  non-uniformity of the kinetic energy of each particle in thermal equilibrium. The formula includes parameters such as mass and temperature.
In addition, one of the models shows  interesting properties where by changing temperature, the non-uniformity varies qualitatively:  outer-energetic at high temperatures and inner-energetic at low temperatures.

This paper is organized as follows.
In Section \ref{sec:equipartition} we briefly review the conventional and generalized principles of equipartition of energy.
 In Section \ref{sec:models} we introduce the proposed model. We outline the  calculation in Section \ref{sec:outline-of-calculation} and the main results in  Section \ref{sec:results}. The final section is devoted to summary and discussion.
 Details of calculation and the Lagrangian the freely jointed chain for an arbitrary number of particles,  are presented in the Appendix.

\section{Principle of Equipartition of Energy}\label{sec:equipartition}
\subsection{principle of equipartition of energy}
Let us briefly summarize the principle of equipartition of energy. 
Suppose we have a classical system described by a Hamiltonian 
\begin{equation}
  \label{eq:hamiltonian-usual}
  H =K(p)+U(q), \ K(p)= \sum_i \frac{1}{2m_i}p_i^2
\end{equation}
where $q=(q_1,q_2,\cdots, q_N)$ and $p=(p_1,p_2,\cdots, p_N)$ are the coordinates and their canonically conjugate momenta, respectively,
$m_i$ are the masses, and  $K(p)$ and $U(q)$ are the kinetic and potential energy of the system, respectively.
If we consider particles in 3-dimension, $N$ is a multiple of 3 and
$m_{i+1}=m_{i+2}=m_{i+3}$.

Let us define the kinetic energy of the $i$'th degree of freedom $K_i$ as
\begin{equation}
  K_i = \frac{1}{2m_i}p_i^2 \ .
\end{equation}
Assume that the system is in thermal equilibrium at temperature $T$.
The principle of equipartition of energy states that the thermal average  of 
the kinetic energy of the  $i$'th degree of freedom
$\left\langle K_i\right\rangle$  is
\begin{equation}
  \label{eq:equipartition-usual}
  \left\langle K_i \right\rangle  = \frac{1}{2}k_B T . 
\end{equation}
Here,  $k_B$ is the Boltzmann constant, the bracket symbol $\left\langle \cdots \right \rangle$ represents thermal average
\begin{equation}
  \left\langle  A\right\rangle = \frac{1}{\mathcal{Z}}\int A e^{-\beta H}dpdq 
\end{equation}
for arbitrary quantity $A$,
$\beta  = 1/k_B T$,   and  $\mathcal{Z}$ is the partition function
\begin{equation}
  \mathcal{Z}=\int e^{-\beta H}dpdq \ .
\end{equation}

Note that the right-hand side of Eq. (\ref{eq:equipartition-usual}) does not depend on any particular degree $i$ of  freedom. Hence,    the average kinetic energy
$\displaystyle \left\langle K_i\right\rangle$ is {\sl uniform} over the system.


The principle also holds when the total kinetic energy  is expressed as
\begin{equation}
  \label{eq:hamiltonian-diag}
  K = \sum_i \frac{1}{2m_i}\alpha_i(q)p_i^2 
\end{equation}
as in a diatomic molecule expressed in polar coordinates.

In the following we introduce the {\sl generalized} principle of equipartition of energy.
In order to distinguish it  from the generalized one, we refer to  the principle of equipartition  as the  ``{\sl conventional principle of equipartition}'' in this paper.

\subsection{generalized principle of equipartition of energy}
Assume that we have a Hamiltonian of the form:
\begin{equation}
  \label{eq:ham-general}
  H = K(q,p) + U(q) \ ,
\end{equation}
where $K(q,p)$ and  $U(q)$ are the kinetic and  potential energies, respectively.
The ``generalized principle of equipartition of energy'' states that, 
if a system whose Hamiltonian is in the form of Eq.(\ref{eq:ham-general})
is in thermal equilibrium at temperature $T$, then
\begin{equation}
  \label{eq:equipartition-general-statement}
  \left\langle \frac{1}{2}p_\xi \frac{\partial H}{\partial p_\xi}\right\rangle = \frac{1}{2}k_B T
\end{equation}
holds for each degree of freedom $\xi$. 
Here, $p_\xi$ represents the canonical momenta, and
repeated indices $\xi$ are not suumed over.
Proof of this principle is similar to that of the conventional case.
Eqs. (\ref{eq:hamiltonian-usual}) and (\ref{eq:hamiltonian-diag}) satisfies   (\ref{eq:equipartition-general-statement}).
Hence, the generalized principle includes the conventional one.

Suppose that the kinetic energy in our Hamiltonian is represented as follows:
\begin{equation}
  K(q,p)=\sum_{\xi,\eta}\frac{1}{2}\alpha_{\xi\eta}(q)p_\xi p_\eta \ . 
\end{equation}
Then
\begin{equation}
\frac{1}{2}p_\xi \frac{\partial H}{\partial p_\xi}   = \sum_{\eta}\frac{1}{2}\alpha_{\xi\eta}(q)p_\xi p_\eta \ne \frac{1}{2m_i}p_i^2 \ .
\end{equation}
Therefore, we observe that
if the off-diagonal elements $\alpha_{\xi\eta}(q)\ne 0$ for $\xi\ne \eta$
the average kinetic energy of each degree of freedom 
may not be constant:
\begin{equation}
  \left\langle\frac{1}{2}m_iv_i^2 \right\rangle \ne \frac{1}{2}k_BT \ . 
\end{equation}
Consequently, the average kinetic energy $\left\langle \frac{1}{2}m_i v_i^2\right\rangle$ depends on $i$. In other words,  it  becomes {\sl non-uniform} under thermal equilibrium.

A class of systems in which the generalized principle of equipartition of energy (\ref{eq:equipartition-general-statement}) holds are systems that have holonomic constraints.
%
%
A well-known example of a system with holonomic constraints is the double pendulum, where the lengths of the links that connect the particles are kept fixed. 
Let $\ell_1$ and $\ell_2$ are links in the double pendulum, and let $m_1$ and $m_2$ are masses
connected to the links, and $\theta_1$ and $\theta_2$ are the angles of the links against the vertically downward direction.
Consequently,  the 
Hamiltonian of the double pendulum is~\cite{Goldstein}
\begin{align}
  H
&=\frac{1}{2\left(m_1+m_2\sin^2(\theta_1-\theta_2)\right)}\nonumber\\
&\left(p_1,p_2\right)
  \begin{pmatrix}
    \frac{1}{\ell_1^2} & -\frac{1}{\ell_1\ell_2}\cos(\theta_1-\theta_2)\\
-\frac{1}{\ell_1\ell_2}\cos(\theta_1-\theta_2) & \frac{m_1+m_2}{m_2}\frac{1}{\ell_2^2}
  \end{pmatrix}
                                                     \begin{pmatrix}
                                                       p_1\\p_2
                                                     \end{pmatrix}
\nonumber\\
&- (m_1+m_2)g\ell_1\cos\theta_1-m_2g\ell_2\cos\theta_2  
  \label{eq:hamiltonian-doublependulum}
\end{align}
We observe the appearance of the off-diagonal elements.

\section{Models}\label{sec:models}
\begin{figure}
  \includegraphics[width=0.4\hsize]{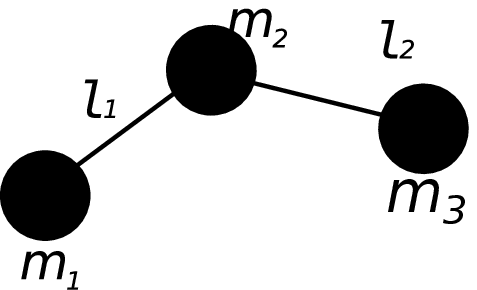}
  \hfill
  \includegraphics[width=0.5\hsize]{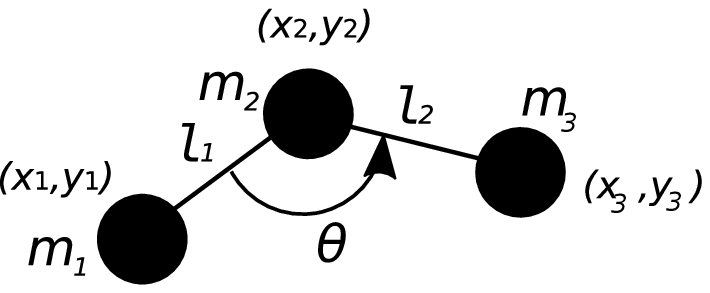}
  \caption{(left) model A : 2-dimensional freely jointed chain with 3 particles \ (right) model B : intra-chain potential is added to model A}
  \label{fig:fjc-2dn3-schem}
\end{figure}

In this study,  we treat two models, one of which is   called a ``freely jointed chain''
\cite{fjc-Kuhn-1934,KUHN194811,Kramers-chain,Fixman3050,doi:10.1063/1.1681834,Mazers-chain-pre-1996,doi-edwards,doi-intro-polymer,strobl-polymer,SiegertWinklerReineker+1993+584+594}
and the other is  its variant. 
They consists of masses serially connected by rigid and light links.
Lagrangians of the freely jointed chain of 2 and 3 dimension with arbitrary number of particles
are shown in Appendix~\ref{sec:app-Lagrangian}. 

The first model, which we refer to  ``model A'', is a 2-dimensional freely jointed chain 
with 3 particles (Fig.~\ref{fig:fjc-2dn3-schem} left).
The second model, which we refer to  ``model B'', is the one which includes the intra-chain
potential in  model A (Fig.~\ref{fig:fjc-2dn3-schem} right).
Let us denote
\begin{equation}
\vec{r}_i=(x_i,y_i)
\end{equation}
 as the coordinate of the $i$'th particle,  
$m_i$ is the mass of the $i$'th particle. 
Both models  A and  B are subject to   the constraint that the distances between adjacent particles are constant.

With this constraint, the Lagrangians of  model A and B are described as follows:
\begin{align}
  \text{model A \ : } &
                          \begin{cases}
                            L &=\sum_{i=1}^3 \frac{m_i}{2} \left(\dot x_i^2+\dot y_i^2\right) \ ,  \\
                            &\left|\vec{r}_{i+1}-\vec{r}_i\right|=\ell_i \ , i=1,2\, .
                          \end{cases}
                              \label{eq:model-A}
  \\
  \text{model B \ : } &
                        \begin{cases}
                          L &=    \sum_{i=1}^3 \frac{m_i}{2} \left(\dot x_i^2+\dot y_i^2\right) -U_0\cos\theta \ , \\ 
                            &\left|\vec{r}_{i+1}-\vec{r}_i\right|=\ell_i \ , i=1,2\, .
                        \end{cases}
                              \label{eq:model-B}
\end{align}
For model B,  $U_0$ is a constant, and  $\theta$ is the angle between the links shown in Fig. \ref{fig:fjc-2dn3-schem}.

The kinetic energy of $i$'th particle $K_i$ in models A and B is expressed as
\begin{equation}
  K_i = \frac{m_i}{2}\left(\dot{x}_i^2+\dot{y}_i^2\right) \ .
\label{eq:k-i}
\end{equation}

Let us  introduce polar coordinate
as
\begin{align}
  x_{i+1}-x_i&=\ell_i \cos\varphi_i  \ , \\
  y_{i+1}-y_i&=\ell_i \sin\varphi_i  \ .
\label{eq:fjc-angles}
\end{align}

Through straight forward calculation, we obtain the kinetic energy 
\begin{align}
  K & = \frac{1}{2}M\left(\dot X_G^2 + \dot Y_G ^2\right) +  \frac{1}{2}{}^t\dot{\vec{\varphi}}\, A\dot{\vec{\varphi}}\ , \\
\vec{\varphi}&=
               \begin{pmatrix}
                 \varphi_1 \\ \varphi_2
               \end{pmatrix}
\label{eq:fjc-2dn3-lagrangian-matrix-1}
\end{align}
where $(X_G, Y_G)$ is the center of mass defined by 
$\displaystyle X_G=\frac{1}{M}\sum_{i=1}^3 m_i x_i$ ,   
$\displaystyle Y_G=\frac{1}{M}\sum_{i=1}^3 m_i y_i$ ,   
and $A$ is  a $2\times 2$ matrix given by
\begin{equation}
  A=M
  \begin{pmatrix}
        \mu_1(\mu_2+\mu_3)\ell_1^2 & \mu_1 \mu_3 \ell_1 \ell_2 \cos\varphi_{12}\\
 \mu_1 \mu_3 \ell_1 \ell_2 \cos\varphi_{12} & \mu_3(\mu_1+\mu_2)\ell_2^2
  \end{pmatrix} \ ,
\label{eq:matrix-a-2d-n3}
\end{equation}
where 
\begin{align}
M&=\sum_{i=1}^3 m_i, \\
\mu_i&=\frac{m_i}{M} \ , \ i=1,2,3 \label{eq:mu-mass}\\
\varphi_{12}&=\varphi_2-\varphi_1=\theta \ .
\end{align}
$\theta$ is the angle appearing in Fig.~\ref{fig:fjc-2dn3-schem}.
From the definition, we have $\sum_{i=1}^3\mu_i=1$ \ .
Using this matrix, the kinetic energy is expressed as
\begin{equation}
  K =\frac{1}{2M}\left|\vec{P_G}\right|^2 + \frac{1}{2}{}^t\vec{p}A^{-1}\vec{p} \ ,
  \label{eq:K-total-momentum}
\end{equation}
where $\vec{P_G} = (P_{G,x},P_{G,y})$ is a momentum of the center of mass, conjugate to
$(X_G, Y_G)$.

Note that  the matrix $A$, and hence, $A^{-1}$ contains off-diagonal elements that depend on the coordinate $\varphi$. This is where the generalized principle of equipartition should be applied to the models A and B.

The Hamiltonian is written as
\begin{align}
  H &= H_G(P,Q) + H_{chain}(p,\varphi) \ ,
      \label{eq:Hamiltonian-total}
  \\
  H_G &= \frac{1}{2M}\left|\vec{P_G}\right|^2 \ , \\
  H_{chain} &= \frac{1}{2}{}^t\vec{p}A^{-1}\vec{p} +U(\varphi) \ , \\
  &
  U(\varphi) =
  \begin{cases}
    0 &\cdots \ \text{model A} \ , \\
    -U_0\cos(\varphi_2-\varphi_1) &\cdots \ \text{model B} \ .
  \end{cases}
\end{align}
where $(P,Q)=(P_{G,x},P_{G,y},X_G,Y_G)$ represent the momenta and coordinates of the center of mass, and
$(p,\varphi)=(p_1,p_2,\varphi_1,\varphi_2)$ represent the momenta and coordinates of the angles of links of the chain.

\section{Outline of Calculation}\label{sec:outline-of-calculation}
Our goal is to obtain the exact expressions for the average kinetic energy of $i$'th particle
at temperature $T$:
\begin{equation}
  \left\langle K_i \right\rangle = \frac{1}{\mathcal{Z}}\int K_i e^{-\beta H}dPdQdpd\varphi \ ,
  \label{eq:Ki-ave-def}
\end{equation}
where
\begin{equation}
  \mathcal{Z}=\int  e^{-\beta H}dPdQdpd\varphi   \ .
\end{equation}
and $\beta = 1/k_B T$. 
Since  the center of mass decouples in the Hamiltonian (\ref{eq:Hamiltonian-total}),
the partition function $\mathcal{Z}$ can be factorized as
\begin{align}
  \mathcal{Z}&=\mathcal{Z}_{G}\mathcal{Z}_{chain}  \ , \\
  \mathcal{Z}_G&=\int  e^{-\beta H_G}dPdQ \ , \\
  \mathcal{Z}_{chain}&=\int  e^{-\beta H_{chain}}dpd\varphi \ ,
\end{align}
and the average $\left\langle K_i\right\rangle$  has the form
  \begin{align}
    \left\langle K_i\right\rangle
    &=
      \frac{\mu_iD}{\beta}+
      \frac{1}{2\beta}\left\langle \mbox{tr}\left(A^{(i)}A^{-1}\right)\right\rangle \
    \\
    &=
            \frac{\mu_i}{\beta}+
      \frac{1}{2\beta}\cdot  \nonumber\\
    &\cdot \frac{1}{\mathcal{Z}_{chain}}\int \mbox{tr}\left(A^{(i)}A^{-1}\right) e^{-\beta H_{chain}}dpd\varphi \  , 
  \end{align}
  where $A^{(i)}$ ($i=1,2,3$) are the $2\times 2$ matrices defined in Eqs.(\ref{eq:a1-2dn3}), (\ref{eq:a2-2dn3}),  and (\ref{eq:a3-2dn3}) in the Appendix~\ref{sec:app-aveKi-A-and-B}.

The integral with respect to $p$ is a 2-dimensional Gaussian integral, and we have
  \begin{align}
    Z_{chain}
&=\sqrt{\frac{2\pi}{\beta}}\int \sqrt{\det A(\varphi)}\,e^{-\beta U(\varphi)} d\varphi \ ,
  \end{align}
\begin{align*}
& 
                 \int
{\rm tr}
\left(
A^{(i)}(\varphi) A^{-1}(\varphi)
                 \right)
                 d\varphi
  \\
&=
 \sqrt{\frac{2\pi }{\beta }}
\int  
{\rm tr}\left(
A^{(i)}(\varphi) A^{-1}(\varphi) 
\right)
\sqrt{\det A(\varphi)} \, e^{-\beta U(\varphi)}
d\varphi \ , 
\end{align*}
where we have used $\det{A^{-1}} = 1/\det A$.
 The details are shown in the Appendix~\ref{sec:app-aveKi}.
Evaluating the integrals with respect to $\varphi$ yields the exact expressions for
$\left\langle K_i\right\rangle$.

\section{Results}\label{sec:results}

\subsection{Freely Jointed Chain (model A)}
\subsubsection{$\left\langle K_i \right\rangle$ for Freely Jointed Chain (model A)}
\begin{figure*}
  \centering
  \includegraphics[width=0.9\hsize]{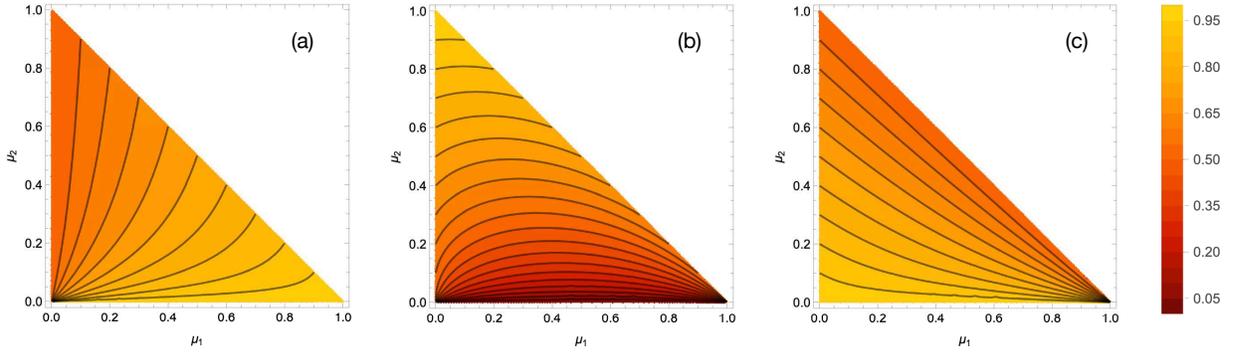}
  \caption{Average kinetic energy per temperature $\left\langle K_i \right\rangle \cdot \beta$ of model A (Eq.(\ref{eq:model-A})).  Panels (a),(b), and (c) represent  $\left\langle K_i\right\rangle$, for $i=1,2,3$, plotted on $(\mu_1,\mu_2)$ plane, respectively. 
}
  \label{fig:ave-Ki-without-potential}
\end{figure*}
Through straightforward calculation we obtain the exact expressions for the thermal average of kinetic energies
for the model at temperature $T$:
\begin{align}
  \left\langle K_1 \right\rangle
  &=
    \frac{1}{\beta}\cdot
    \left\{
    \mu_1 
    +
    (\mu_2+\mu_3)
    \left(
    1-
    \frac{1}{2}
    \frac{\mu_2
    }
    {
    \mu_2+\mu_1\mu_3
    }
    \frac{
    K(k)
    }
    {
    E(k)
    }
    \right)
\right\}
\label{eq:ave-linear-ki1-n3d2}
  \end{align}
\begin{align}
     \left\langle K_2 \right\rangle
  &=
    \frac{1}{\beta}
    \left\{
\mu_2 
+
\mu_2
\left(
\frac{(1+\mu_2)}{2}
\frac{1
}
{
\mu_2+\mu_1\mu_3
}
\frac{
K(k)
}
{
E(k)
}
    -1
    \right)
\right\} 
    \label{eq:ave-linear-ki2-n3d2}
    %
\end{align}
\begin{align}
     \left\langle K_3 \right\rangle
  &=
    \frac{1}{\beta}\cdot
    \left\{
\mu_3 
+
(\mu_1+\mu_2)
\left(
1
-
\frac{1  }{2}
\frac{
\mu_2}
{
\mu_2+\mu_1\mu_3
}
\frac{
K(k)
}
{
E(k)
    }
    \right)
\right\}
\label{eq:ave-linear-ki3-n3d2} 
\end{align}
where  $\mu_i$ is defined in Eq.(\ref{eq:mu-mass}), 
\begin{align}
  k^2&\equiv \frac{\mu_1\mu_3}{\mu_2+\mu_1\mu_3}  =\frac{\mu_1\mu_3}{(1-\mu_1)(1-\mu_3)}
       \label{eq:parameter-k2}
\end{align}
and $K(k)$ and $E(k)$ are complete elliptic integrals of the first and the second kinds, respectively~\cite{DLMF}:
\begin{align}
  K(k)&\equiv \int_0^{\frac{\pi}{2}}\left(1-k^2(\sin\theta)^2\right)^{-\frac{1}{2}}d\theta \ ,\\
  E(k)&\equiv \int_0^{\frac{\pi}{2}}\left(
1-k^2 (\sin\theta)^2
\right)^{\frac{1}{2}}d\theta \ , 
\end{align}
We observe that $\sum_{i=1}^3 \left\langle K_i\right\rangle = 2/\beta$, as expected.
Details are shown in Appendix~\ref{sec:app-aveKi-A-and-B}.

From Eqs.(\ref{eq:ave-linear-ki1-n3d2}),(\ref{eq:ave-linear-ki2-n3d2}) and
(\ref{eq:ave-linear-ki3-n3d2}), we observe that  $\left\langle K_i\right \rangle$
depends on $\beta $ and $\mu_i$, and not on $\ell_i$.

Since for model A  the dependence of $\left\langle K_i\right \rangle$ on $\beta$ is
\begin{equation}
  \left\langle K_i\right \rangle \propto \frac{1}{\beta}
\end{equation}
and $\mu_1+\mu_2+\mu_3=1$, we can plot the result
(\ref{eq:ave-linear-ki1-n3d2}) and  (\ref{eq:ave-linear-ki2-n3d2})  and
(\ref{eq:ave-linear-ki3-n3d2}) on the  parameter plane $(\mu_1,\mu_2)$
as Fig.~\ref{fig:ave-Ki-without-potential}.
It is clear that the average kinetic energy $\left\langle K_i\right\rangle$ 
is non-uniform.

\begin{figure}
  \centering
  \includegraphics[width=\hsize]{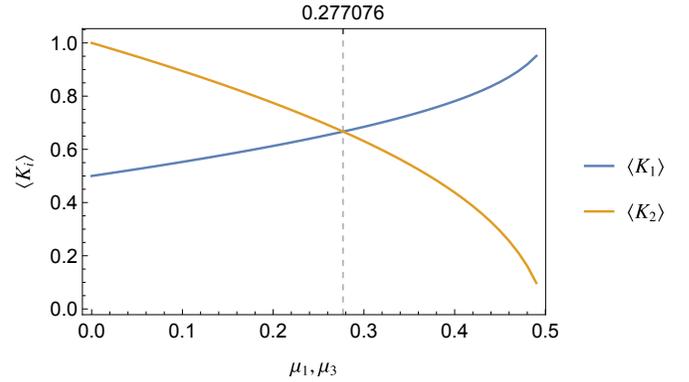}
  \caption{model A : $\left\langle K_1 \right\rangle$ and $\left\langle K_2 \right\rangle$ vs $\mu_1$ when $m_1=m_3$. The vertical broken line indicates the value of $\mu_1=\mu_1^*$ where $\left\langle K_1 \right\rangle = \left\langle K_2 \right\rangle$ is shown at the top of the figure. $\beta=1$.}
  \label{fig:ki-linear-vs-m1-for-2dn3-free}
\end{figure}

In order to grasp the result we present a simple case where the mass distribution is symmetric: $m_1 = m_3 $. Consequently,  $\mu_1 = \mu_3$ and $\left\langle K_1\right\rangle =\left\langle K_3\right\rangle$.
The results are shown in Fig.~\ref{fig:ki-linear-vs-m1-for-2dn3-free}.
Roughly speaking, $\left\langle K_1\right\rangle$ (and $\left\langle K_3\right\rangle$)
is large when $\mu_1=\mu_3$ is large, and vice versa. Moreover we observe that
the order of magnitude among $\left\langle K_i\right\rangle$ varies
when varying $\mu_1$. 
That is, 
 we can control the order of 
magnitude of $ \left\langle K_i \right\rangle$ ($i=1,2,3$)  by appropriately choosing the value of $\mu_1$.

Initially  we note that
$
  \left\langle K_1 \right\rangle
=  \left\langle K_2 \right\rangle 
$
at a value of  $\mu_1 $ which satisfies
\begin{equation}
  1-\frac{3}{2}\frac{1-2\mu_1}{1-\mu_1}\frac{K(k)}{E(k)} =0\ ,
  \label{eq:mu1-special}
\end{equation}
here $k^2$ is determined from $\mu_1$ as
\begin{equation}
  k^2 = \frac{\mu_1\mu_3}{\mu_2+\mu_1\mu_3}=\left(\frac{\mu_1}{1-\mu_1}\right)^2 \ .
\end{equation}
by eq. (\ref{eq:parameter-k2}).

Let us denote the value $\mu_1=\mu_1^*$ which satisfies Eq.(\ref{eq:mu1-special}).
We note that
\begin{itemize}
\item when $\mu_1<\mu_1^*$\, , 
the inner particle has more average kinetic energy
$\left\langle K_1 \right\rangle =\left\langle K_3 \right\rangle < \left\langle K_2 \right\rangle$ .
\item when $\mu_1>\mu_1^*$\, , 
the outer particles have more average kinetic energy
$\left\langle K_1 \right\rangle =\left\langle K_3 \right\rangle > \left\langle K_2 \right\rangle$ .
\end{itemize}

We obtained the value of $\mu_1^*$ numerically and it is shown in
Fig.~\ref{fig:ki-linear-vs-m1-for-2dn3-free}.
\subsubsection{$\left\langle K_i^2 \right\rangle$ for Freely Jointed Chain (model A)}
\begin{figure*}
  \centering
  \includegraphics[width=0.9\hsize]{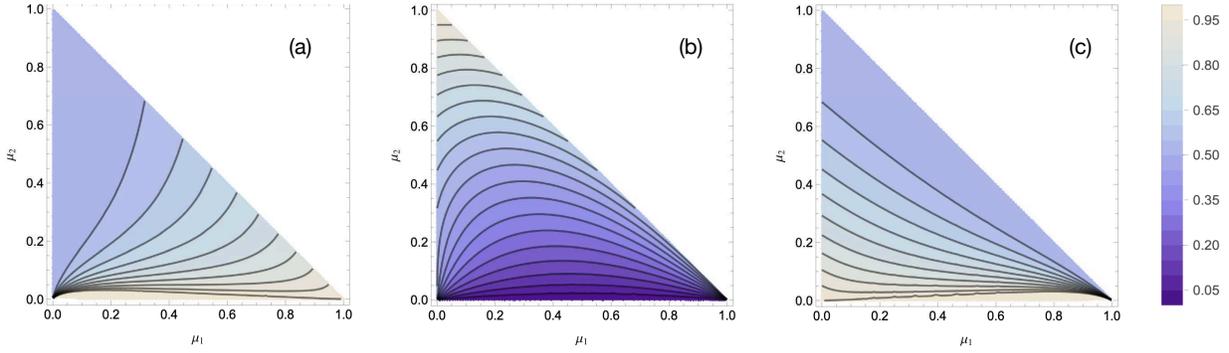}
  \caption{
Variances of  kinetic energy  per squared temperature $\left\langle (\Delta K_i)^2\right\rangle\cdot \beta^2 =\left\langle \left(K_i-\left\langle K_i\right\rangle\right)^2\right\rangle \cdot \beta^2$ of model A (Eq.(\ref{eq:model-A})).  Panels (a),(b) and (c) represent $\left\langle (\Delta K_i)^2\right\rangle$   for $i=1,2,3$, plotted on the $(\mu_1,\mu_2)$ plane, respectively. 
}
  \label{fig:ave-Ki2-without-potential}
\end{figure*}
For this model, we can also obtain exact expressions of $\left\langle K_i^2 \right\rangle$
and hence $\left\langle (\Delta K_i)^2 \right\rangle =\left\langle \left(K_i-\left\langle K_i \right\rangle\right)^2 \right\rangle  = \left\langle K_i^2 \right\rangle - \left(\left\langle K_i \right\rangle\right)^2$, which represent
local fluctuation of kinetic energy.

The exact expressions of $\left\langle K_i^2 \right\rangle$ are  lengthy. Hence
we put them in the Appendix \ref{sec:k2-a}, and we demonstrate the $\left\langle (\Delta K_i)^2 \right\rangle$ in Fig.~\ref{fig:ave-Ki2-without-potential} on the $(\mu_1,\mu_2)$ plane. 
Since both $\left\langle K_i^2 \right\rangle$ and $\left\langle K_i^2 \right\rangle$ are
proportional to $\beta^{-2}$ we set $\beta=1$ in the figure without loss of generality.

\begin{figure}
  \centering
  \includegraphics[width=\hsize]{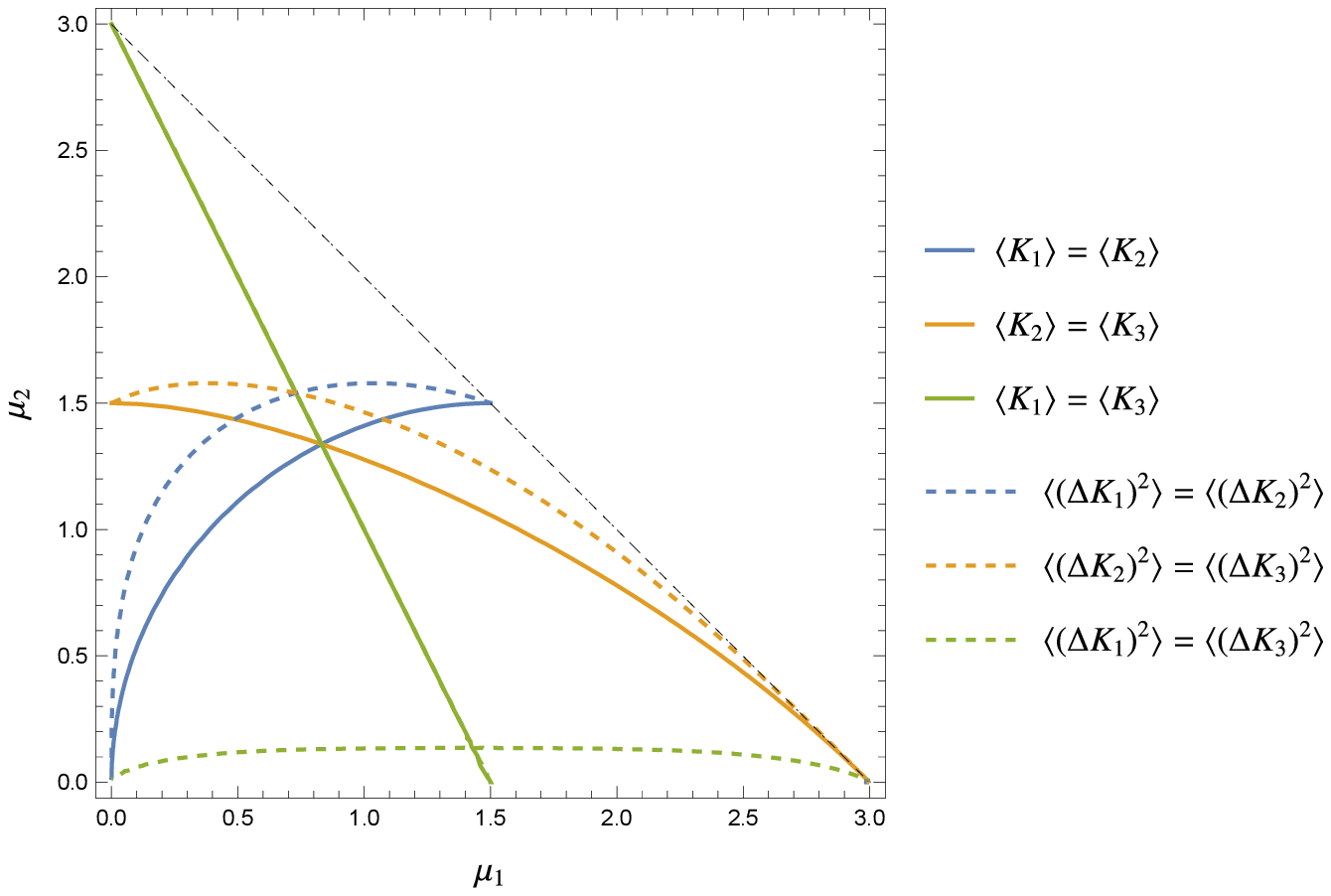}
  \caption{$\displaystyle \left\langle K_i\right\rangle=\displaystyle \left\langle K_j\right\rangle$ (solid line) and $\displaystyle \left\langle (\Delta K_i)^2\right\rangle=\displaystyle \left\langle (\Delta K_j^2)\right\rangle$ (dashed line) shown on a parameter space $(\mu_1,\mu_2)$ for model A. }
  \label{fig:ave-Ki-Kj-Ki2-Kj2-fjc}
\end{figure}
In Fig.\ref{fig:ave-Ki-Kj-Ki2-Kj2-fjc} we demonstrate, on the $(m_1,m_2)$ plane the boundaries
$\left\langle K_i\right\rangle = \left\langle K_j\right\rangle$ as solid lines
and
$\left\langle (\Delta K_i)^2\right\rangle = \left\langle (\Delta K_j)^2\right\rangle$
as dashed lines.
Note that the solid  and dashed lines are not identical.
Thus we observe that for there are some parameter regions  $(m_1,m_2)$ and  choice of
particles $(i,j)$
the order of magnitude is inverted for
$\left\langle K_i\right\rangle $
and
$\left\langle (\Delta K_i)^2\right\rangle$, 
that is,
$\left\langle K_i\right\rangle > \left\langle K_j\right\rangle$
and
$\left\langle (\Delta K_i)^2\right\rangle < \left\langle (\Delta K_j)^2\right\rangle$
holds.

As shown  above,  the parameters where $\left\langle (\Delta K_i)^2\right\rangle = \left\langle (\Delta K_j)^2\right\rangle$ differ from the parameters
where
$\left\langle  K_i\right\rangle = \left\langle  K_j\right\rangle$
is quite different from the systems of conventional
Hamiltonian Eq.(\ref{eq:hamiltonian-usual}), where
$\left\langle K_i\right\rangle = \frac{1}{2\beta}$, $\left\langle K_i^2\right\rangle = \frac{3}{4\beta^2}$, $\left\langle \left(\Delta K_i\right)^2\right\rangle=\frac{1}{2\beta^2}$ and the variance of kinetic energy is  uniform and proportional to the square of the average kinetic energy.
\subsection{Freely Jointed Chain with Intra-Chain Potential (model B)}
\subsubsection{$\left\langle K_i \right\rangle$ for Freely Jointed Chain with Intra-Chain Potential (model B)}
\begin{figure*}
  \centering
  \includegraphics[width=0.9\hsize]{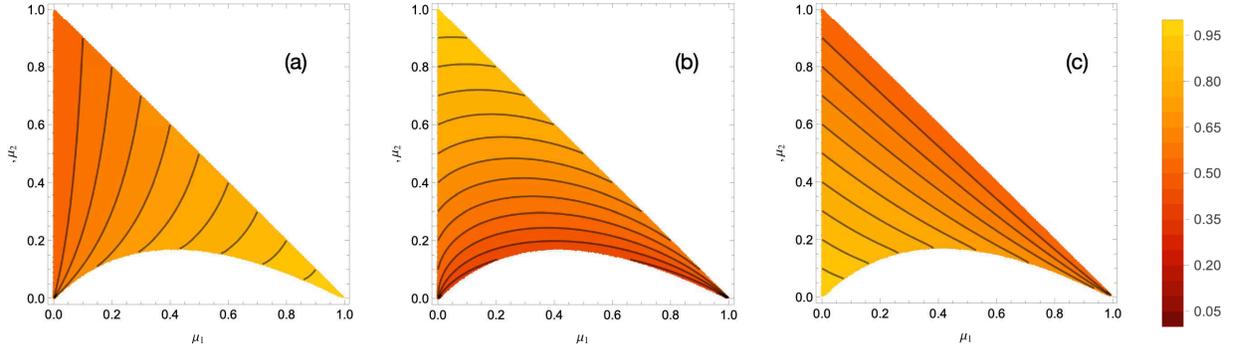}
  \caption{average kinetic energy $\left\langle K_i \right\rangle$ of model B (Eq.(\ref{eq:model-B})).  Panels (a),(b) and (c) represent  $\left\langle K_i\right\rangle$ for $i=1,2,3$, plotted on the $(\mu_1,\mu_2)$ plane, respectively. 
Regions where $\mu_2$ is small are shown blank because the series expansion of the expression is not valid in this region.
$\beta=1$, $U_0=1$ .
}
  \label{fig:ave-Ki-with-potential}
\end{figure*}

Let us examine model B (Eq.(\ref{eq:model-B})),
where there is a potential with respect to the angle between the links.
When $\tilde{\mu}\equiv\frac{\mu_1\mu_3}{\mu_2}<1$ we obtain $\left\langle K_i \right\rangle$ by series expansion.
The results are as follows:
\begin{align}
  \left\langle K_i\right\rangle &=\frac{1}{\beta}\left(\mu_i+F_i \right)
\label{eq:ave-linear-ki-for-intra-pot}
\end{align}
where
\begin{align}
F_i &= \frac{1}{2}\left\langle \mbox{tr}\left(A^{(i)}A^{-1}\right)\right\rangle \, \\
F_1 &=(\mu_2+\mu_3)\left\{1-\frac{1}{2}\frac{W(1)}{W(3)}     \right\}\\
F_2 &= (1+\mu_2) \frac{1}{2}\frac{W(1)}{W(3)}-\mu_2
      \\
F_3 &=(\mu_1+\mu_2)\left\{ 1-\frac{1}{2}\frac{W(1)}{W(3)}
      \right\} 
\end{align}
and $W(c)$ is defined by the following series
\begin{align}
W(c)&=I_0(\beta U_0) \nonumber\\
  &+ \frac{1}{\sqrt{\pi}} \sum_{n=1}^\infty (-1)^n \frac{(2n-c)!!}{n!}{\tilde \mu}^n 
\frac{\Gamma(n+\frac{1}{2})}{(\beta U_0)^n} I_n(\beta U_0), \\
& \phantom{+ \frac{1}{\sqrt{\pi}\sum_{n=1}^\infty (-1)^n \frac{(2n-c)!!}{n!}}} \ \text{where} \ \tilde \mu=\frac{\mu_1\mu_3}{\mu_2} \ .
\label{eq:W-def}
\end{align}
and $I_n(z)$ is the  modified Bessel function of the $n$-th order~\cite{DLMF}.
The above expression is valid for $0\leq \tilde\mu <1$.
We note that $F_1+F_2+F_3=1$ and $\sum_{i=1}^3\left\langle K_i\right\rangle = 2/\beta$, as expected.
Details are shown in Appendix~\ref{sec:app-aveKi-A-and-B}.

Note that the function $W(c)$ depends on $\beta$. Hence 
the distribution of kinetic energy changes upon varying the temperature.
The result is shown in the parameter space $(\mu_1,\mu_2)$ in Fig.~\ref{fig:ave-Ki-with-potential}.


\subsubsection{Temperature dependence and inside-outside crossover}
\begin{figure}
  \centering
    \includegraphics[width=0.8\hsize]{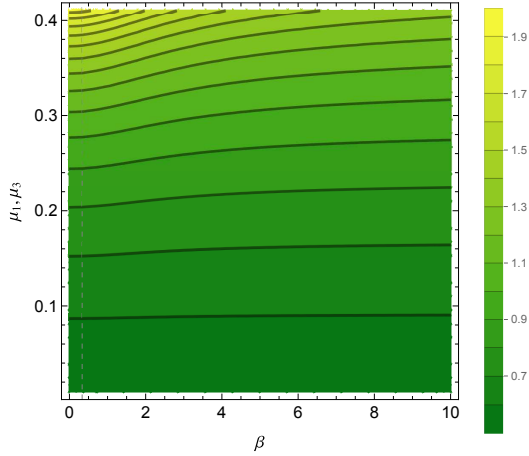}
    \caption{$\frac{\left\langle K_1 \right\rangle}{\left\langle K_2 \right\rangle}$  vs $\beta$ and $m_1$ for model B (Eq.(\ref{eq:model-B})). $m_1=m_3$. The horizontal and vertical axes represent $\beta$ and $m_1$, respectively. $U_0=1$.}
    \label{fig:ki-ave-linear-with-pot-sym}
\end{figure}

To grasp the above result, let us consider a simple case in which the  mass distribution is symmetric : $m_1 = m_3$.  Consequently,  $\mu_1=\mu_3$ and 
$\left\langle K_1 \right\rangle =\left\langle K_3 \right\rangle$. 
Hence, we measure the distribution of kinetic energy 
by the ratio
$\displaystyle 
\frac{\left\langle K_1 \right\rangle}{\left\langle K_2 \right\rangle} 
$, which is shown in 
Fig.~\ref{fig:ki-ave-linear-with-pot-sym}.
This figure demonstrates that, in addition to the mass distribution, $\left\langle K_i \right\rangle$  depends on the temperature.

\begin{figure}
  \centering
    \includegraphics[width=0.7\hsize]{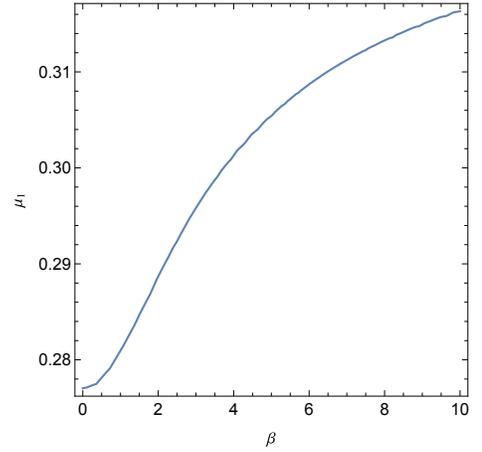}
    \caption{Boundary where $\left\langle K_1 \right\rangle  =\left\langle K_2 \right\rangle = \left\langle K_3 \right\rangle$    plotted on $(\beta,\mu_1)$ plane  for model B (Eq.(\ref{eq:model-B})). 
$\left\langle K_1 \right\rangle  =\left\langle K_3 \right\rangle < \left\langle K_2 \right\rangle$ for the right of the boundary, and $\left\langle K_1 \right\rangle =\left\langle K_3 \right\rangle  > \left\langle K_2 \right\rangle$ for the left of the boundary. $U_0=1$ . }
    \label{fig:ki-ave-linear-with-pot-sym-bd}
\end{figure}
And for some range of $\mu_1$, there is a temperature where
all the average kinetic energy is uniform:
$
\left\langle K_1 \right\rangle =\left\langle K_3 \right\rangle = \left\langle K_2 \right\rangle 
$\ .
We show the boundary in $(\beta,\mu_1)$-parameter space where $
\left\langle K_1 \right\rangle =\left\langle K_3 \right\rangle = \left\langle K_2 \right\rangle $
holds in Fig.~\ref{fig:ki-ave-linear-with-pot-sym-bd}.

Let us denote the temperature as $T=T^*$ where
$\left\langle K_1 \right\rangle =\left\langle K_3 \right\rangle = \left\langle K_2 \right\rangle$
holds.
We note  that even in fixed mass distribution, by changing temperature 
\begin{itemize}
\item when $T<T^*$;  the inner particle has more average kinetic energy:
  $\left\langle K_1 \right\rangle =\left\langle K_3 \right\rangle < \left\langle K_2 \right\rangle$ \, , 
  
\item when $T>T^*$; the outer particles have more average kinetic energy:
  $\left\langle K_1 \right\rangle =\left\langle K_3 \right\rangle > \left\langle K_2 \right\rangle$ \, .
\end{itemize}

\begin{figure*}
  \centering
   \includegraphics[width=0.45\hsize]{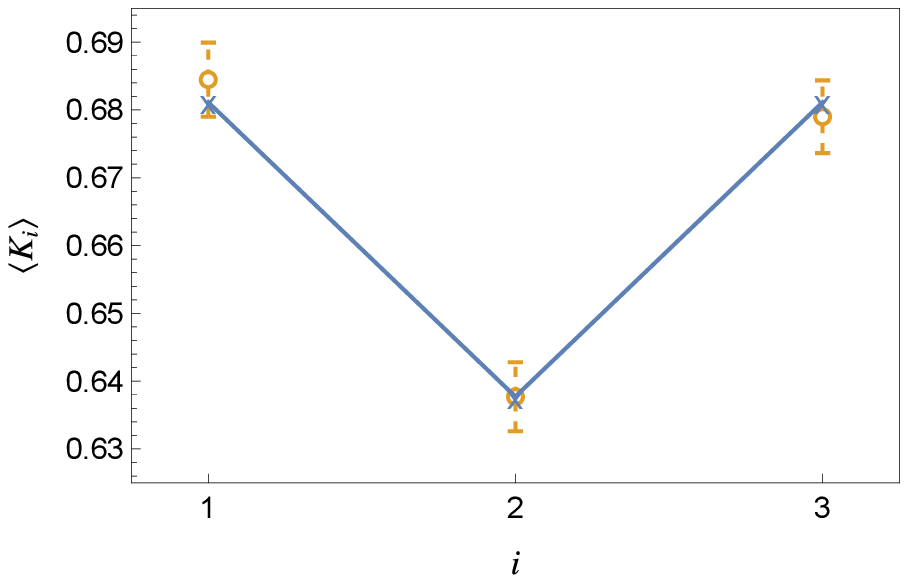}
   \includegraphics[width=0.45\hsize]{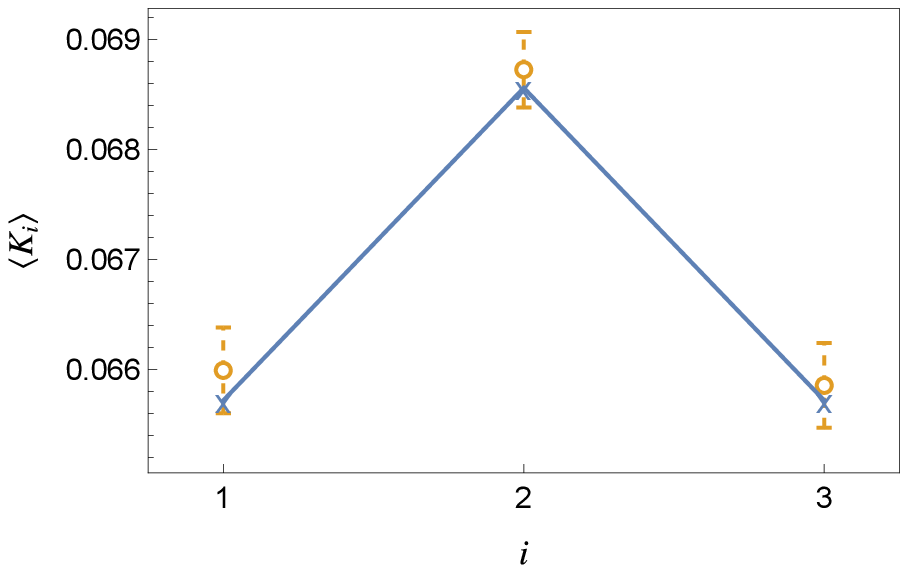}
  \caption{Difference of  kinetic energy distribution by temperature for the model B (Eq.(\ref{eq:model-B})). The horizontal axis represents the index $i$ (position) of the particles,  and the vertical axis represent the average kinetic energy $\displaystyle \left\langle K_i \right\rangle$. The solid lines are  connecting the theoretical values (\ref{eq:ave-linear-ki-for-intra-pot}), and symbols with error bars represent values obtained by the MCMC simulation based on Eq.(\ref{eq:Ki-ave-def}).  $\mu_1=\mu_3=0.3$, $\mu_2=0.4$.  $U_0=1.0$. At high temperature, the  outer particles have more average kinetic energy (left figure,  $\beta=1$) , and at low temperature, the inner particle has a higher  average kinetic energy. (right figure, $\beta=10$)}
  \label{fig:ave-linear-ki-theory-and-mcmc}
\end{figure*}

This result is  confirmed by numerically computing $\displaystyle \left\langle K_i\right\rangle$ using the Markov chain Monte Carlo (MCMC) method. Fig.~\ref{fig:ave-linear-ki-theory-and-mcmc}  shows $\displaystyle \left\langle K_i\right\rangle$ for $\mu_1=\mu_3=0.3$, $\mu_2=0.4$, where we can note that the values of  $\displaystyle \left\langle K_i\right\rangle$ computed by MCMC agree well with theoretical values in  Eq.(\ref{eq:ave-linear-ki-for-intra-pot}), and the profile of the average kinetic energy $\left\langle K_i\right\rangle$ varies
upon changing the temperature.

\subsubsection{$\left\langle (K_i)^2 \right\rangle$ for Freely Jointed Chain with Intra-Chain Potential (model B)}
\begin{figure*}
  \centering
   \includegraphics[width=0.9\hsize]{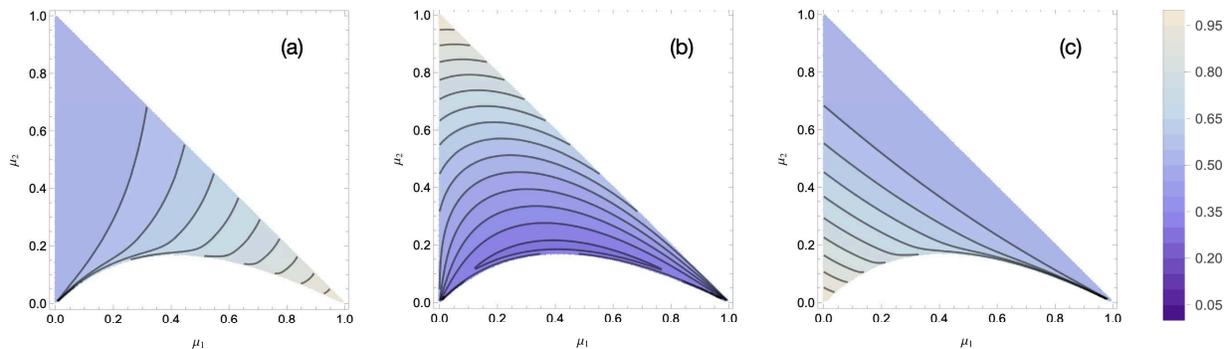}
  \caption{Variances of  kinetic energy $\left\langle (\Delta K_i)^2\right\rangle=\left\langle \left(K_i-\left\langle K_i\right\rangle\right)^2\right\rangle $ of model B (\ref{eq:model-B}).  Panels (a),(b) and (c) represent $\left\langle (\Delta K_i)^2\right\rangle$   for $i=1,2,3$, plotted on the $(\mu_1,\mu_2)$ plane, respectively. 
Regions where $\mu_2$ is small are shown blank because the series expansion of the expression is not valid in this region.
$\beta=1$, $U_0=1$
}
  \label{fig:ave-Ki2-with-potential}
\end{figure*}

For model B also we are able to obtain the exact expressions of
$\left\langle (K_i)^2 \right\rangle$.
The exact form is  presented  in the Appendix~\ref{sec:k2-b}. Using this we can calculate 
the local fluctuation of the kinetic energy $\left\langle (\Delta K_i)^2\right\rangle$
in Fig.~\ref{fig:ave-Ki2-with-potential}.

\begin{figure}
  \centering
  \includegraphics[width=\hsize]{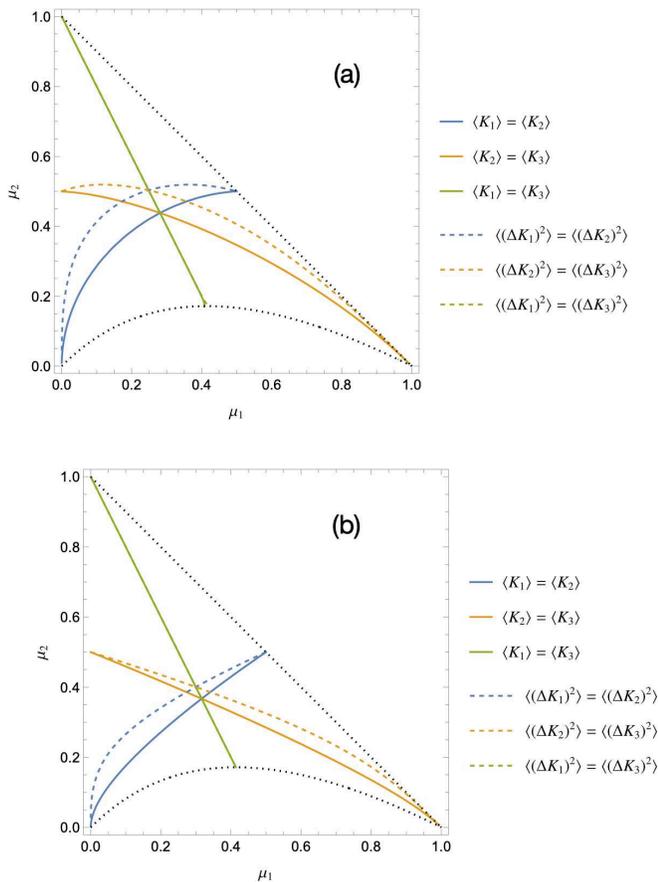}
  \caption{$\displaystyle \left\langle K_i\right\rangle=\displaystyle \left\langle K_j\right\rangle$ (solid line) and $\displaystyle \left\langle (\Delta K_i)^2\right\rangle=\displaystyle \left\langle (\Delta K_j^2)\right\rangle$ (dashed line) shown on a parameter space $(\mu_1,\mu_2)$ for model B. $U_0$=1. Panels (a) and  (b) represent  (a)$\beta=1$ and  (b)$\beta=10$, respectively.  }

  \label{fig:ave-Ki-Kj-Ki2-Kj2-fjc-with-potential}
\end{figure}

 In Fig.\ref{fig:ave-Ki-Kj-Ki2-Kj2-fjc-with-potential} we show on a $(\mu_1,\mu_2)$ space the boundary  $\left\langle K_i\right\rangle =\left\langle K_j\right\rangle$
 ($i,j=1,2,3$) and $\left\langle \left(\Delta K_i\right)^2\right\rangle =\left\langle\left(\Delta  K_j\right)^2\right\rangle$ for $\beta=1$ and $\beta=10$.
The boundaries move by changing the temperature, hence we can control the order $\left\langle K_i \right\rangle > \left\langle K_j \right\rangle$ and $\left\langle \left(\Delta K_i \right)^2 \right\rangle >\left\langle \left(\Delta K_i \right)^2 \right\rangle $ by changing the temperature while keeping the values of mass fixed.


\section{Summary and Discussions}
In this paper we show the exact expressions of each particle's thermal averages of kinetic energy in 3-particle 2-dimensional freely jointed chain and a related model. 
The results clearly confirm the spontaneous emergence of non-uniformity in kinetic energy in thermal equilibrium. This is a direct result of the generalized principles of equipartition of energy.
Previously, the non-uniformity of average kinetic energy was obtained by
 numerical and approximate analytic methods~\cite{yanagita-gakkai-1,yanagita-gakkai-2,yanagita-gakkai-3,chain-letter-JSTAT-2009,chain-JSTAT-2010}. This is the first result
where the non-uniform average kinetic energy $\left\langle K_i \right\rangle$ is exactly
obtained.

In addition,  using the analytic expression we found that the profile  of the average kinetic energy varies qualitatively upon changing parameters such as  the mass ratio and the  coefficient of  potentials in relative angles.
In particular, for the model with intra-angle potential, we found that,  by changing the temperature only, the most energetic particles switch from outermost particles to the inner one. This result shows the possibility of controlling the activity in many-body systems with constraints.

An example  that can be related  to this result is the concept of ``local temperature''. It is  useful to utilize this  concept in biological systems ~\cite{Vale-Oosawa-1990-AdvBP}. It will be interesting to examine the relationship  between the non-uniformity of the temperature  in biological systems and that  of the  average  kinetic energy found in this paper.

We also show the explicit form of the thermal average of the  variance of the kinetic energy of each
particle.
Contrary to  the case with the  usual Hamiltonian ( Eq.(\ref{eq:hamiltonian-usual})),  the variance is not proportional to the square of the average kinetic energy.  
The fact that the variance is not always proportional to the
squared average is particularly interesting if we consider the variance as the  ``local  specific heat'' .

In this study,  all the models  considered have rigid constraints. It is known that 
there are no exact rigid bodies in real world  and that rigid bodies in the model systems are
idealizations of stiff potentials or hard springs, which admit the conventional principle of 
equipartition of energy. We note that, if the potential is quite stiff,  therelaxation time for
conventional equipartition is quite long. The relaxation times are estimated by using Boltzmann-Jeans theory~\cite{konishi-yanagita-chain-2,konishi-yanagita-chain-3} and  are typically  exponentially long with respect to  the frequency of oscillation in the potential, and we  have a good chance of observing the non-uniformity.

It will be important to consider how the non-uniformity discussed in this study is realized in real-world systems. Candidates are proteins, which are large chain type systems.
Proteins usually perform proper functions in a properly folded form. However, certain proteins referred to as ``intrinsically disordered protein''~\cite{doi:10.1021/cr400514h} do not take a fixed form,  and some parts of the chain are not folded.  Examples are found that   such proteins  interact with other proteins via the unfolded part ~\cite{WRIGHT1999321}.
In such cases, the unfolded part can have a larger kinetic energy and moves actively, and consequently,  play an important role in searching for another molecule for the reaction.
The activity of the unfolded part can be related to the property studied in this study where
the particles at the end of chain have a large average kinetic energy.

It will also be interesting to consider the process of protein folding, where an elongated chain is gradually folded over time. Let us consider the coupling between molecules as constraint. The first constraint is the primary structure of the protein chain. In the course of folding some molecules make strong connections, which become newly introduced constraints.
By introducing such constraints, the non-uniformity of the average kinetic energy can vary.

\begin{acknowledgements}
    T. K. is supported by Chubu University Grant (A).
T.Y. acknowledges the support of Japan Society for the Promotion of Science (JSPS) KAKENHI Grants No. 18K03471 and 21K03411.
  \end{acknowledgements}



  \appendix
  \section{Derivation of Lagrangian of the models}\label{sec:app-Lagrangian}
  \subsection{Derivation of Lagrangian of Freely Jointed Chain and variant for  arbitrary length $N$ in 2 and 3 dimension}\label{sec:app-fjc-2d3d-N}
  Here, we first derive the Lagrangian of freely jointed chain and its variant that is  composed of an arbitrary number  of particles.  The models are in two  and three spatial dimensions.

  The model is composed of particles that are serially connected by massless rigid links. Let us put an
  index for each particle in accordance with the order of their position from one end of the chain to the other.
  Let us denote $\vec{r}_i$, $i=1,2,\cdots, N$, as the position of the $i$-th particle,  $m_i$ as the mass of
  the $i$-th particle, $\ell_i$ as the length of the $i$-th link that  connects particles $i$ and $i+1$,
  $U(r)$ as the potential function (which vanishes for the original freely jointed chain).
Consequently, the Lagrangian (with constraints) is written as
  \begin{align}
      L &= K(\dot r) - U(r) \ , \\
    K&=\sum_{i=1}^{N-1}\frac{1}{2}m_i\left|\dot{\vec{r_i}}\right|^2,
    \label{eq:kinetic-original}\\
        & \left|\vec{r}_{i+1}-\vec{r}_i\right|=\ell_i \ , i=1,2, \cdots, N-1.
  \end{align}
where  $\vec{r}_i = (x_i,y_i)$ for 2-dimensions and   $\vec{r}_i = (x_i,y_i,z_i)$ for 3-dimensions.
 The potential function $U(q)$ is absent for the original freely jointed chain.
 For later purposes,  we denote $M$ as the total mass and $\mu_i$ as the relative mass of
 the $i$-th particle:
 \begin{equation}
   M = \sum_{i=1}^N m_i \ , \ \
   \mu_i = \frac{m_i}{M} \ .
 \end{equation}
 Note that $\sum_{i=1}^N \mu_i=1$ holds.

Let us write $\vec{s}_i$ as the unit vector that is  directed from $\vec{r}_i$ to $\vec{r}_{i+1}$.
Then
\begin{equation}
  \vec{r}_{i+1}=\vec{r}_i +\ell_i \vec{s_i}  \ \ \ ( i=1,2,\cdots,N-1) \ .
  \label{eq:qi-successive}
\end{equation}

We express coordinates of $\vec{s_i}$ as
\begin{equation}
\vec{s_i} \equiv
  \begin{pmatrix}
 \cos\varphi_i \\
     \sin\varphi_i 
  \end{pmatrix}
\end{equation}
in 2-dimension  and 
\begin{equation}
\vec{s_i} \equiv
  \begin{pmatrix}
    \sin\theta_i \cos\varphi_i \\
    \sin\theta_i \sin\varphi_i \\
    \cos\theta_i
  \end{pmatrix}
\end{equation}
in 3-dimension.

The center of mass is defined as
  \begin{equation}
    \vec{Q_G} \equiv \frac{1}{M}\sum_{i=1}^N m_i \vec{r}_i=\sum_{i=1}^N \mu_i \vec{r}_i \ .
    \label{eq:center-of-mass-def}
  \end{equation}

The  kinetic energy is expressed  in terms of the center of mass $\vec{Q_G}$  and the directional unit vector $\vec{s_i}$.
From Eq.(\ref{eq:qi-successive}) we note that all  $\vec{r}_i$'s ($i=2,3,\cdots,N$) are
expressed in terms of $\vec{r}_1$ and $\vec{s_i}$:
\begin{equation}
  \vec{r}_i = \vec{r}_1 +   \sum_{j=1}^{i-1}\ell_j \vec{s_j} \ .
  \label{eq:ri-r1-si}
\end{equation}

$\vec{Q_G}$ is expressed in terms of $\vec{q}_1$ and $\vec{s_i}$, $i=1,2,\cdots,N-1$.
Consequently,  we note that all $\vec{r}_i$ , $i=1,2,\cdots, N$ are expressed in terms of 
$\vec{Q}_G$ and  $\vec{s_i}$, $i=1,2,\cdots,N-1$ as
\begin{equation}
  \vec{r}_i = \vec{Q}_G + \sum_{j=1}^{N-1} a_{ij} \vec{s_j} \ ,
\label{eq:ri-and-rg-and-sj-aij}
\end{equation}
where the coefficients $a_{ij}$ are defined as
\begin{equation}
  a_{ij}\equiv
  \begin{cases}
    \phantom{-}\mu_j^{\le}\,\ell_j & \cdots j < i \\
    -\mu_j^{>}\, \ell_j &\cdots j \ge i . 
  \end{cases}
  \label{eq:def-aij}
\end{equation}
and  the coefficients $\mu^{\le}_i$ ,  $\mu^{>}_i$,  are 
are defined as follows:
\begin{equation}
  \mu^{\le}_i \equiv \sum_{k=1}^i \mu_k \ \ , \ \ 
  \mu^{>}_i \equiv \sum_{k=i+1}^N \mu_k  \ .
\end{equation}

Consequently, in terms of $\vec{Q_G}$ and $\vec{s_i}$ the kinetic energy (Eq.(\ref{eq:kinetic-original}))
is expressed as follows:
\begin{equation}
  K = \frac{1}{2}M\left| \dot{\vec{Q_G}} \right|^2
+
\frac{1}{2}M\sum_{j=1}^{N-1} \sum_{k=1}^{N-1} \mu^{\le}_{\min(j,k)}\mu^{>}_{\max(j,k)}
\ell_j\ell_k\dot{\protect\vec{s_j}}\cdot\dot{\protect\vec{s_k}} \ . 
\label{eq:ke-total}
\end{equation}


For the 2-dimensional model,
\begin{align}
  K &= \frac{1}{2}M\left| \dot{\vec{Q_G}} \right|^2\nonumber\\
&+
\frac{1}{2}M\sum_{j=1}^{N-1} \sum_{k=1}^{N-1} \mu^{\le}_{\min(j,k)}\mu^{>}_{\max(j,k)}
\ell_j\ell_k\cos(\varphi_j-\varphi_k) \dot{\varphi_j}\dot{\varphi_k}\ , 
\label{eq:ke-total-2dim}
\end{align}
and for the 3-dimensional one,
\begin{align}
  K &= \frac{1}{2}M\left| \dot{\vec{Q_G}} \right|^2
+
\frac{1}{2}M\sum_{j=1}^{N-1} \sum_{k=1}^{N-1} \mu^{\le}_{\min(j,k)}\mu^{>}_{\max(j,k)}
\ell_j\ell_k\cdot \nonumber\\
    & \cdot
      \left\{
      \dot{\theta_j}\dot{\theta_k}\cos\theta_j\cos\theta_k\cos(\varphi_j-\varphi_k)
      \right.
      \nonumber\\
    &+
      \dot{\theta_j}\dot{\varphi_k}\cos\theta_j\sin\theta_k\sin(\varphi_j-\varphi_k)\nonumber\\
    &-
      \dot{\varphi_j}\dot{\theta_k}\sin\theta_j\cos\theta_k\sin(\varphi_j-\varphi_k)\nonumber\\
    &+
      \dot{\varphi_j}\dot{\varphi_k}\sin\theta_j\sin\theta_k\cos(\varphi_j-\varphi_k)\nonumber\\
    &+
      \left.
      \dot{\theta_j}\dot{\theta_k}\sin\theta_j\sin\theta_k 
  \right\} \ .
\label{eq:ke-total-3dim}
\end{align}
Hence, the Lagrangian of a 2-dimensional freely jointed chain is
\begin{align}
  L &= \frac{1}{2}M\left| \dot{\vec{Q_G}} \right|^2\nonumber\\
&+
\frac{1}{2}M\sum_{j=1}^{N-1} \sum_{k=1}^{N-1} \mu^{\le}_{\min(j,k)}\mu^{>}_{\max(j,k)}
\ell_j\ell_k\cos(\varphi_j-\varphi_k) \dot{\varphi_j}\dot{\varphi_k}\ , 
\label{eq:Lagrangian-total-2dim}
\end{align}
and the same for a  3-dimensional freely jointed chain is
\begin{align}
  L &= \frac{1}{2}M\left| \dot{\vec{Q_G}} \right|^2
+
\frac{1}{2}M\sum_{j=1}^{N-1} \sum_{k=1}^{N-1} \mu^{\le}_{\min(j,k)}\mu^{>}_{\max(j,k)}
\ell_j\ell_k\cdot \nonumber\\
    & \cdot
      \left\{
      \dot{\theta_j}\dot{\theta_k}\cos\theta_j\cos\theta_k\cos(\varphi_j-\varphi_k)
      \right.
      \nonumber\\
    &+
      \dot{\theta_j}\dot{\varphi_k}\cos\theta_j\sin\theta_k\sin(\varphi_j-\varphi_k)\nonumber\\
    &-
      \dot{\varphi_j}\dot{\theta_k}\sin\theta_j\cos\theta_k\sin(\varphi_j-\varphi_k)\nonumber\\
    &+
      \dot{\varphi_j}\dot{\varphi_k}\sin\theta_j\sin\theta_k\cos(\varphi_j-\varphi_k)\nonumber\\
    &+
      \left.
      \dot{\theta_j}\dot{\theta_k}\sin\theta_j\sin\theta_k 
  \right\} \ .
\label{eq:Lagrangian-total-3dim}
\end{align}

\subsection{Lagrangian for model A and model B}
Model A (Eq.(\ref{eq:model-A})) and model B (Eq.(\ref{eq:model-B})) are defined in
two-dimensions and have 3 particles.

Setting $N=3$ in Eq.(\ref{eq:ke-total-2dim}), we have
\begin{equation}
  K = \frac{1}{2}M\left(\dot{X_G}^2+\dot{Y_G}^2\right)
      +\frac{1}{2}{}^t\dot{\vec{\varphi}}A\dot{\vec{\varphi} }\ ,
      \label{eq:k-total-in-velocity}
    \end{equation}  
where
\begin{align}
  A &=
      M
  \begin{pmatrix}
        \mu_1(\mu_2+\mu_3)\ell_1^2 & \mu_1 \mu_3 \ell_1 \ell_2 \cos\varphi_{12}\\
 \mu_1 \mu_3 \ell_1 \ell_2 \cos\varphi_{12} & \mu_3(\mu_1+\mu_2)\ell_2^2
\end{pmatrix} \ ,
\label{eq:matA-2dim-N3}
\\
  \vec{\varphi} &=
                  \begin{pmatrix}
                    \varphi_1 \\ \varphi_2
                  \end{pmatrix}
  \ ,\\
  \varphi_{12}&=\varphi_2-\varphi_1 \ .
\end{align}

Using this kinetic energy, the Lagrangian of model A and model B are
\begin{equation}
  L = K - U(q) \ , 
\end{equation}
where $K$ is defined in Eq.(\ref{eq:k-total-in-velocity}) and 
\begin{equation}
  U(q) =
  \begin{cases}
    0 & \cdots \ \text{model A} \ , \\
    -U_0 \cos(\varphi_2-\varphi_1 )& \cdots \ \text{model B} \ .
  \end{cases}
\end{equation}

\section{Derivation of $\left\langle K_i \right\rangle$}\label{sec:app-aveKi}
\subsection{preparation}
Let us denote the angular variable as $q$. That is, $q=\varphi$ in two dimension
and $q=(\theta,\varphi)$ in three dimensions.
Remember that $N$ is the number of particles. Hence, 
\begin{align}
  q&=\left(\varphi_1,\cdots,\varphi_{N-1}\right) \ \text{in 2-dim.} \ , \\
  q   &=\left(\theta_1,\varphi_1,\theta_2,\varphi_2,\cdots,\theta_{N-1},\varphi_{N-1}\right) 
\ \text{in 3-dim}.
\end{align}
Let us denote the number of angular coordinate as $\nu$:
\begin{equation}
  \nu = 
  \begin{cases}
    N-1 & \cdots \ \text{in 2-dim.} \ \\
   2(N-1) & \cdots \ \text{in 3-dim.} \ \\
  \end{cases}
  \label{eq:number-of-angular-variables}
\end{equation}
In addition to $q$, we have the coordinates of the center of mass 
as $(X_G,Y_G)$ in two dimensions and $(X_G,Y_G,Z_G)$ in three dimensions.
Hence, the total number of degrees of freedom is
\begin{equation}
\begin{cases}
  2+(N-1) &= N+1   \ \text{in 2-dim.} \ , \\
  3+2(N-1) &= 2N+1  \ \text{in 3-dim.} 
\end{cases}
\end{equation}

Through straightforward calculation, the kinetic energy of the particle $i$ is
\begin{align}
  K_i &= \frac{1}{2}M\mu_i
\dot{Q}_G^2
 + \frac{1}{2}{}^t\dot{q}A^{(i)}\dot{q}
 \nonumber\\
        & + \ \text{(cross terms between components of  $\dot Q_G$ and $\dot q$)} \ ,
\end{align}
where $A^{(i)}$s are square matrices depending on $q$.

Let us denote $P_G$ and $p$ as momenta conjugate to $Q_G$ and $q$, respectively.
In terms of the momenta, we have
\begin{align}
  K_i &= \frac{1}{2M}\mu_i\dot{P}_G^2 + \frac{1}{2}{}^tpA^{-1}A^{(i)}A^{-1}p \nonumber\\
       & + \ \text{(cross terms between components of  $P_G$ and $p$)} \ ,
         \label{eq:Ki-linear-in-momenta-general}
\end{align}

Since the cross terms vanish on integration with respect to  components of $P_G$ or $p$, we have
\begin{align}
  \left\langle K_i \right\rangle
&=\mu_i\cdot\frac{D}{2\beta} 
+\frac{1}{2}\left\langle
{}^tpA^{-1}A^{(i)}A^{-1}p
\right\rangle \ .
\end{align}
Here, we have used $\left\langle K_i \right\rangle=\frac{D}{2\beta}$ and $D$ as the number of
spatial dimensions. 
Integration by parts in the second term yields
\begin{align}
  \left\langle K_i \right\rangle
&=\mu_i\cdot\frac{D}{2\beta} 
+\frac{1}{2\beta}\left\langle
\mbox{tr}\left(A^{(i)}A^{-1}\right)
\right\rangle
\label{eq: Ki-ave-tr-general}
\end{align}


Since the motion of the center of mass decouples, as shown in Eq.(\ref{eq:Hamiltonian-total}),
we observe that
\begin{align}
  \mathcal{Z}&=\mathcal{Z}_{G}\mathcal{Z}_{chain}  \ , \\
  \mathcal{Z}_G&=\int  e^{-\beta H_G}dP_GdQ_G \ , \\
  \mathcal{Z}_{chain}&=\int  e^{-\beta H_{chain}}dpdq \ .
\end{align}

Note that the Hamiltonian has the form of Eq.(\ref{eq:Hamiltonian-total}): 
we first integrate by $p$. The integration is essentially a $\nu$-dimensional Gaussian integral:
\begin{equation}
  \int  \exp\left(-\beta\frac{1}{2}pA^{-1}p\right)dp
=\left(\frac{2\pi}{\beta}\right)^{\nu/2}\sqrt{ \det A}
\end{equation}
where we have used $\det{A^{-1}} = 1/\det A$ and $\nu$ is defined in Eq.(\ref{eq:number-of-angular-variables}).

Using this, we obtain the partition function for angular degrees of freedom $Z_{chain}$ as
\begin{align}
  Z_{chain}&\equiv \int e^{-\beta\left(\frac{1}{2}pA^{-1}p + U(q)\right)}dpdq
\nonumber\\
&=\left(\frac{2\pi}{\beta}\right)^{\nu/2}
\int \sqrt{\det A}\,e^{-\beta U(q)} dq \ .
\label{eq:Z-chain}
\end{align}

\begin{align}
& 
\left\langle
{\rm tr}
\left(
A^{(i)} A^{-1}
\right)
\right\rangle  
=
\left(\frac{2\pi}{\beta}\right)^{\nu/2}
\frac{1}{Z_{chain}}
 \nonumber
\\
&
\cdot \int  
{\rm tr}\left(
A^{(i)} A^{-1}
\right)
\sqrt{\det A} \, e^{-\beta U(q)}
dq
\label{eq:tr-ave}
\end{align}

By calculating Eqs.(\ref{eq:Z-chain}), (\ref{eq:tr-ave}) and substituting them into Eq.(\ref{eq: Ki-ave-tr-general})  we obtain expressions for
$\left\langle K_i \right\rangle$.

\subsection{Model A and model B, $N=3$ and 2-dimensional case}\label{sec:app-aveKi-A-and-B}
Let us turn our attention from the case of  general degrees of freedom and the general spatial dimension to  the case of model A (Eq.(\ref{eq:model-A})) and model B (Eq.(\ref{eq:model-B})) , that is, to the 2-dimensional 3-particle case ($N=3$).
 From Eq.(\ref{eq:number-of-angular-variables}) the number of angular coordinate is $\nu = N-1=2$.

The $2\times 2$ matrices $A^{(i)}$ appearing in Eq.(\ref{eq:tr-ave}) are
\begin{align}
A^{(1)} & =M\mu_1
    \nonumber\\
& \cdot\begin{pmatrix}
    (\mu_2+\mu_3)^2\ell_1^2 & \mu_3 (\mu_2+\mu_3)\ell_1 \ell_2 C_{12}\\
\mu_3 (\mu_2+\mu_3)\ell_1 \ell_2 C_{12} &\mu_3^2 \ell_2^2 
  \end{pmatrix}
\label{eq:a1-2dn3}\\
&{}\nonumber\\
A^{(2)}
&=
          M\mu_2
          \begin{pmatrix}
    \mu_1^2 \ell_1^2 & -\mu_1\mu_3\ell_1\ell_2C_{12}\\
-\mu_1\mu_3\ell_1\ell_2C_{12} & \mu_3^2\ell_2^2
  \end{pmatrix}
\label{eq:a2-2dn3}\\
&{}\nonumber\\
A^{(3)}
&=
          M\mu_3
          \nonumber\\
& \cdot \begin{pmatrix}
\mu_1^2\ell_1^2 & \mu_1(\mu_1+\mu_2)\ell_1\ell_2C_{12}\\
\mu_1(\mu_1+\mu_2)\ell_1\ell_2C_{12} &(\mu_1+\mu_2)^2\ell_2^2
  \end{pmatrix}
\label{eq:a3-2dn3}
\end{align}
where
\begin{equation}
  C_{12}=\cos(\varphi_2-\varphi_1) \ .
\end{equation}

From Eq.(\ref{eq:matrix-a-2d-n3}) we have
\begin{equation}
 \det A 
=
M^2 \ell_1^2\ell_2^2\mu_1\mu_2\mu_3\left\{
1+\frac{\mu_1\mu_3}{\mu_2}\left(\sin\varphi_{12})\right)^2
\right\}  \ .
\label{eq:deta-d2n3}
\end{equation}


The matrix $A$ in Eq.(\ref{eq:tr-ave})  is shown in Eq.(\ref{eq:matA-2dim-N3}).

  If we denote
\begin{equation}
  F\equiv \mu_2\left\{1+\frac{\mu_1\mu_3}{\mu_2}\left(\sin\varphi_{12}\right)^2\right\} \, , 
\label{eq:F}
\end{equation}
one can express ${\rm tr}\left(A^{(i)}(q) A^{-1}(q) \right)$,  $i=1,2,3$, as
\begin{align}
{\rm tr}\left(A^{(1)}A^{-1}\right)
&=  \frac{(\mu_2+\mu_3)}{F}(2F-\mu_2) 
\label{eq:trA1Ainv-in-F}
\\
{\rm tr}\left(A^{(2)}A^{-1}\right)
&=  \frac{\mu_2}{F}(1+\mu_2-2F) 
\label{eq:trA2Ainv-in-F}
\\
{\rm tr}\left(A^{(3)}A^{-1}\right)
&=  \frac{(\mu_1+\mu_2)}{F}(2F-\mu_2) 
\label{eq:trA3Ainv-in-F}
\end{align}
and consequently, we have
\begin{align}
&  {\rm tr}\left(A^{(1)}A^{-1}\right)
+{\rm tr}\left(A^{(2)}A^{-1}\right)
+{\rm tr}\left(A^{(3)}A^{-1}\right) 
\nonumber\\
&=2 \ ,
\end{align}
which yields
\begin{equation}
     \left\langle K_1 \right\rangle
+     \left\langle K_2 \right\rangle
+     \left\langle K_3 \right\rangle
=
\frac{1}{\beta} + \frac{1}{2\beta}\times 2 = \frac{2}{\beta} \ ,
\end{equation}
as expected.

Substituting Eqs.(\ref{eq:trA1Ainv-in-F}),  (\ref{eq:trA2Ainv-in-F}) and (\ref{eq:trA3Ainv-in-F}) into Eq.(\ref{eq:tr-ave})  and evaluating the integral, we arrive at  the exact 
expressions for $\left\langle K_i \right\rangle$ shown in Sec. \ref{sec:results}.

For model A (Eq.(\ref{eq:model-A})), setting $U(q)=0$ in Eq.(\ref{eq:tr-ave}),
we observe that the average $\left\langle K_i \right\rangle$ is expressed by 
complete elliptic integrals. The final results are shown in
 Eqs.(\ref{eq:ave-linear-ki1-n3d2}) , (\ref{eq:ave-linear-ki2-n3d2}) and (\ref{eq:ave-linear-ki3-n3d2}).

 For model B (Eq.(\ref{eq:model-B})),
we expand $\det A$  in power series of $\displaystyle \tilde{\mu}\sin^2\varphi_{12}$,
where $\tilde{\mu}=\frac{\mu_1\mu_3}{\mu_2} $ \ .
Consequently, each term of the expansion is proportional to $\int \sin^{2\nu}\theta \,e^{\beta U_0\cos\theta}d\theta$ for some integer $\nu$. After some manipulations the integral is expressed in terms of modified Bessel function $I_\nu(\beta U_0)$, and the final result  for $\left\langle K_i \right\rangle$ is shown in Eqs.(\ref{eq:ave-linear-ki-for-intra-pot}) $\sim$ (\ref{eq:W-def}). 

\section{Derivation of $\left\langle K_i^2 \right\rangle$}
\subsection{Derivation for both model A and B}
Let us restrict ourselves to the 2-dimensional model with $N=3$.
Using  Eq.(\ref{eq:Ki-linear-in-momenta-general}), the kinetic energy of the particle $i$,  is explicitly written as:
\begin{align}
  K_i
%
&=
 \mu_i K_G
+ \frac{\mu_i}{2M}\sum_{\alpha,\gamma,\zeta,\eta=1,2}\left(A^{(i)}_{\alpha \gamma}A^{-1}_{\alpha\zeta}A^{-1}_{\gamma,\eta}\right)\,p_\zeta \,p_{\eta} \nonumber\\
&+\sum_{\xi=x,y}\sum_{\eta=1,2}G^{(i)}_{\xi,\eta}\,P_\xi\, p_\eta  
\label{eq:Ki-linear}
\ , 
\end{align}
where
\begin{align}
  K_G &=\frac{1}{2M}\left( P_{x} ^2+P_{y}^2 \right)\\
  G^{(i)}_{x,\eta}&=\frac{\mu_i}{M}\sum_{\alpha=1,2}a_{i,\alpha}\cos\varphi_\alpha A^{-1}_{\alpha,\eta}
\label{eq:def-G-x}
\\
  G^{(i)}_{y,\eta}&=\frac{\mu_i}{M}\sum_{\alpha=1,2}a_{i,\alpha}\sin\varphi_\alpha A^{-1}_{\alpha,\eta}
\label{eq:def-G-y}
\end{align}
and 
\begin{equation}
  a_{ij}
=
  \begin{pmatrix}
    -(\mu_2+\mu_3)\ell_1 & -\mu_3 \ell_2 \\
\mu_1 \ell_1 &-\mu_3\ell_2\\
\mu_1\ell_1 & (\mu_1+\mu_2)\ell_2
  \end{pmatrix} \ .
\label{eq:aij-2dn3}
\end{equation}

Through straightforward calculation we obtain
\begin{align}
  \left\langle
  \left(K_i\right)^2  
\right\rangle
&=
                2\frac{\mu_i}{\beta}\left\langle K_i \right\rangle 
                \nonumber\\
&+
  \left\langle
 \left(\frac{\mu_i}{2M}\sum_{\alpha,\gamma,\zeta,\eta=1,2}\left(A^{(i)}_{\alpha \gamma}A^{-1}_{\alpha\zeta}A^{-1}_{\gamma,\eta}\right)\,p_\zeta \,p_{\eta}\right)^2
\right\rangle 
\nonumber\\
&+
  \left\langle
\left(\sum_{\xi=x,y}\sum_{\eta=1,2}G^{(i)}_{\xi,\eta}\,P_\xi\, p_\eta  \right)^2 
\right\rangle .
\label{eq:ki2-temp}
\end{align}

Using integration by parts,  the second term in Eq.(\ref{eq:ki2-temp}) can be expressed as follows: 
\begin{align}
&     \left\langle
\left(  \sum_{\alpha,\gamma,\zeta,\eta=1,2}
\left(A^{(i)}_{\alpha \gamma}A^{-1}_{\alpha\zeta}A^{-1}_{\gamma,\eta}\right)\,p_\zeta \,p_{\eta}\right)^2
  \right\rangle
  \nonumber\\
&=\frac{2M^2}{\beta^2}
\left\langle
\mbox{tr}\left\{\left(A^{(i)}A^{-1}\right)^2 \right\}
\right\rangle
+  
\frac{M^2}{\beta^2}
\left\langle
\left(\mbox{tr}\left\{A^{(i)}A^{-1}\right\}\right)^2
\right\rangle \ ,
\label{eq:AApppp-partial-4}
\end{align}

Again, through a By straightforward calculation the third term of Eq.(\ref{eq:ki2-temp}) is expressed 
as follows:
\begin{align}
&\left\langle\left(\sum_{\xi=x,y}\sum_{\eta=1,2}G^{(i)}_{\xi,\eta}\,P_\xi\, p_\eta  \right)^2 \right\rangle
=\mu_i^2\frac{1}{\beta^2}
\left\{
  a_{i,1}^2\left\langle
 A^{-1}_{1,1}
\right\rangle 
\right.
\nonumber\\
&
\phantom{\mu_i^2\frac{1}{\beta^2}}
\left.
+
    a_{i,2}^2\left\langle
A^{-1}_{2,2}
\right\rangle 
+
2
  a_{i,1} a_{i,2}\left\langle
  \cos\left(\varphi_2-\varphi_1\right)
A^{-1}_{1,2}
\right\rangle 
\right\}
\label{eq:GPp-sq-ave-2}
\end{align}

Using the equations obtained above,  the averages of the squared kinetic energy
$\left\langle K_i^2 \right\rangle$ in models A and B are expressed as follows:
\begin{description}
\item[$i=1$ \ \ ] 
\begin{align}
  \left\langle
  \left(K_1\right)^2  
\right\rangle
&=
2\frac{\mu_1}{\beta}\left\langle K_1 \right\rangle 
\nonumber\\
&+
\frac{1}{4\beta^2}
\left(\mu_2+\mu_3\right)^2
\left\{
6
-
4\mu_2\left\langle\frac{1}{F}\right\rangle
\right.
\nonumber\\
&
\left.
+
\mu_2^2\left\langle\frac{1}{F^2}\right\rangle
+
2\mu_1^2\mu_3^2\left\langle\frac{S^4}{F^2}\right\rangle
\right\}
\nonumber\\
&+
\frac{\mu_1(\mu_2+\mu_3)}{\beta^2}
\left\{
2-\mu_2
\left\langle\frac{1}{F}\right\rangle 
\right\}
\label{eq:k12-temp2}
\end{align}

\item[$i=2$ \  \ ] 
\begin{align}
  \left\langle
  \left(K_2\right)^2  
\right\rangle
&=
2\frac{\mu_2}{\beta}\left\langle K_2 \right\rangle 
\nonumber\\
&+\frac{\mu_2^2}{4}\frac{1}{\beta^2}
\left\{
\left[2(\mu_1+\mu_3)^2+(1+\mu_2)^2\right]
\left\langle\frac{1}{F^2}\right\rangle
\right. \nonumber\\
&
\left. -4(1+\mu_2)
\left\langle\frac{1}{F}\right\rangle
-4\mu_1\mu_3\left[
1+(\mu_1+\mu_3)
\right]\left\langle \frac{S^2}{F^2}\right\rangle
\right. \nonumber\\
&
\left.
+4+4(\mu_1+\mu_3)\mu_1^2\mu_3^2\left\langle \frac{S^4}{F^2}\right\rangle
\right\}
 \nonumber\\
&
+
\frac{\mu_2^2}{\beta^2}
\left\{
\left(1+\mu_2\right)
\left\langle
\frac{1}{F}
\right\rangle 
-2
\right\}
\label{eq:k22-temp2}
\end{align}

\item[$i=3$ \  \ ] 
\begin{align}
  \left\langle
  \left(K_3\right)^2  
\right\rangle
&=
2\frac{\mu_3}{\beta}\left\langle K_3 \right\rangle
\nonumber\\
&+
\frac{1}{4\beta^2}(\mu_1+\mu_2)^2
\left\{
6-4\mu_2\left\langle\frac{1}{F}\right\rangle
\right. \nonumber\\
&\left.
 + \mu_2^2\left\langle\frac{1}{F^2}\right\rangle
+2\mu_1^2\mu_3^2\left\langle\frac{S^4}{F^2}\right\rangle
\right\}
\nonumber\\
&+
\frac{\mu_3(\mu_1+\mu_2)}{\beta^2}
\left\{
2-\mu_2
\left\langle
\frac{1}{F}
\right\rangle 
\right\}
\label{eq:k32-temp2}
\end{align}
\end{description}
Here $F$ is defined in Eq.(\ref{eq:F}) and 
\begin{equation}
  S = \sin\theta \ .
\end{equation}

The outcome  differs from model A and model B because of the potential
$-U_0 \cos\varphi_{12}$.
\subsection{Averages of functions appearing in the exact expressions of $\left\langle K_i^2 \right\rangle$ for model A (Eq.(\ref{eq:model-A}))}\label{sec:k2-a}
For model A (Eq.(\ref{eq:model-A})), the averages 
$\left\langle K_i^2 \right\rangle$ (Eqs.(\ref{eq:k12-temp2}), (\ref{eq:k22-temp2}) and (\ref{eq:k32-temp2})) are obtained as follows:
\begin{align}
  \left\langle \frac{1}{F}\right\rangle
%
%
&=
  \frac{1}{\mu_2+\mu_1\mu_3}
\frac{K(k)}{E(k)} \ , 
\label{eq:ave-1-over-F}
\end{align}

\begin{align}
  \left\langle \frac{1}{F^2}\right\rangle 
%
&=
\frac{1}{\left(\mu_2 + \mu_1\mu_3\right)^{2}}
  \frac{1}{  E(k)}
\Pi(k^2,k)
%
\end{align}

\begin{align}
  \left\langle \frac{S^2}{F^2}\right\rangle 
&=
\frac{(k^2-1)\Pi(k^2,k)+K(k)}{\left(\mu_2 + \mu_1\mu_3\right)^{2}k^2\,E(k)}
\label{eq:ave-S2-over-F2}
\end{align}
\begin{align}
  \left\langle \frac{S^4}{F^2}\right\rangle 
%
&=
  \frac{(k^2-1)^2\Pi(k^2,k)
+2(k^2-1)K(k) +E(k)}{\left(\mu_2 + \mu_1\mu_3\right)^{5/2}k^4 E(k)}
\end{align}

Here, $\Pi(\alpha,k^2)$ is the complete elliptic integral of the third kind,  and is defined as~\cite{DLMF}
\begin{equation}
  \Pi(\alpha^2,k) = \int_0^{\pi/2}\frac{d\theta}{ (1-\alpha^2\sin^2\theta)\sqrt{1-k^2\sin^2\theta\, {}}} \ .
  \label{eq:elliptic-Pi}
\end{equation}

Since $\left\langle K_i\right\rangle \propto 1/\beta$, we note that 
\begin{equation}
  \left\langle \left(\Delta K_i\right)^2\right\rangle \propto 1/\beta^2 \ .
\end{equation}
\subsection{Averages of functions appearing in the exact expressions of $\left\langle K_i^2 \right\rangle$ for model B (Eq.(\ref{eq:model-B}))}\label{sec:k2-b}
For model B (Eq.(\ref{eq:model-B})), the averages
$\left\langle K_i^2 \right\rangle$ (Eqs.(\ref{eq:k12-temp2}), (\ref{eq:k22-temp2}) and (\ref{eq:k32-temp2})) are obtained as follows:
\begin{align}
  \left\langle \frac{1}{F}\right\rangle
&=
\frac{\mu_2^{-1}}{Z_2}\cdot \sum_{n=0}^\infty\frac{1}{n!}\cdot \frac{(-1)^n\left[(2n-1)!!\right]^2}{2^n}\cdot \tilde{\mu}^n \frac{I_n(\beta U_0)}{(\beta U_0)^n}
\end{align}
\begin{align}
  \left\langle \frac{1}{F^2}\right\rangle 
&=
\frac{\mu_2^{-2}}{Z_2}\cdot \sum_{n=0}^\infty\frac{1}{n!}\cdot \frac{(-1)^n(2n+1)\left[(2n-1)!!\right]^2}{2^n}\nonumber\\
&\cdot \tilde{\mu}^n \frac{I_n(\beta U_0)}{(\beta U_0)^n}
\end{align}

\begin{align}
  \left\langle \frac{S^2}{F^2}\right\rangle 
&=
\frac{\mu_2^{-2}}{Z_2}\cdot \sum_{n=0}^\infty\frac{1}{n!}\cdot \frac{(-1)^n\left[(2n+1)!!\right]^2}{2^n}\cdot \tilde{\mu}^n \frac{I_{n+1}(\beta U_0)}{(\beta U_0)^{n+1}}
\end{align}
\begin{align}
  \left\langle \frac{S^4}{F^2}\right\rangle 
&=
    \frac{\mu_2^{-2}}{Z_2}\cdot \sum_{n=0}^\infty\frac{1}{n!}\cdot \frac{(-1)^n(2n+3)\left[(2n+1)!!\right]^2}{2^n}
    \nonumber\\
  &
    \phantom{ \frac{\mu_2^{-2}}{Z_2}\cdot \sum_{n=0}^\infty
    }
    \cdot\tilde{\mu}^n \frac{I_{n+2}(\beta U_0)}{(\beta U_0)^{n+2}} \ , 
\end{align}
And
\begin{align}
Z_2 &\equiv
 I_0(\beta U_0) 
 +\sum_{n=1}^\infty\frac{1}{n!}\cdot  \frac{(-1)^{n-1}(2n-1)\left[(2n-3)!!\right]^2}{2^n}
 \nonumber\\
 &
 \phantom{ I_0(\beta U_0) 
 +\sum_{n=1}^\infty
   }
\cdot \tilde{\mu}^n
\frac{ I_n(\beta U_0) }{\left(\beta U_0\right)^n} \ . 
\end{align}
Here $I_n(z)$ is the modified Bessel function of the $n$-th order~\cite{DLMF}.

%

\end{document}